\documentclass[sigconf]{acmart}
\AtBeginDocument{%
  }

\usepackage{siunitx}
\usepackage{pgf}
\usepackage{pgfmath}
\usepackage{xparse}
\usepackage{siunitx}
\usepackage{float}
\usepackage{dblfloatfix}
\usepackage{placeins}   

\newcommand{\signif}[1]{%
    \pgfmathparse{#1}%
    \ifdim \pgfmathresult pt < 0.0001pt %
        $\boldsymbol{\pgfmathprintnumber[assume math mode=true]{\pgfmathresult}}$***
    \else
        \ifdim \pgfmathresult pt < 0.001pt %
            $\boldsymbol{\pgfmathprintnumber[assume math mode=true]{\pgfmathresult}}$**
        \else
            \ifdim \pgfmathresult pt < 0.005pt %
                $\boldsymbol{\pgfmathprintnumber[assume math mode=true]{\pgfmathresult}}$*
            \else
                $\pgfmathprintnumber[assume math mode=true]{\pgfmathresult}$
            \fi
        \fi
    \fi
}

\usepackage{subcaption}
\usepackage{xcolor,colortbl}
\usepackage[edges]{forest}
\usepackage{amsmath}
\usepackage{siunitx}
\usepackage{enumitem}
\usepackage{graphicx}
\graphicspath{{figures/}}

\newcommand{\pl}{PrairieLearn{}~\citep{west2015prairielearn}}

\newcommand{\uci}{State Research University~\#2}
\newcommand{\usu}{State Research University~\#1}

\begin{document}

\title{Assessing Student Ability to Select an Algorithmic Paradigm}

\author{Dip Kiran Pradhan Newar}
\orcid{0009-0003-3499-3645}
\affiliation{%
  \institution{Utah State University}
  \city{Utah}
  \country{USA}}
\email{dip.pradhannewar@usu.edu}

\author{Michael Shindler}
\orcid{0000-0002-3365-1729}
\affiliation{%
  \institution{University of California, Irvine}
  \city{CA}
  \country{USA}}
\email{mikes@uci.edu}

\author{Seth Poulsen}
\orcid{0000-0001-6284-9972}
\affiliation{%
  \institution{Utah State University}
  \city{Utah}
  \country{USA}}
\email{seth.poulsen@usu.edu}
\renewcommand{\shortauthors}{Dip Kiran Pradhan Newar, Michael Shindler, \& Seth Poulsen}




\begin{abstract}
Computer science students are expected to be able to look at a problem and select an appropriate algorithm design paradigm to use to produce a solution. However, there is little research on how students determine which algorithmic paradigm to use. Historically, researchers have relied on free-response questions or interviews to assess students' knowledge of algorithmic paradigm selection. To successfully evaluate and scale teaching interventions for selecting an algorithmic design paradigm, we need to efficiently test a student's ability to select among different design paradigms. Here, we present the first attempts to assess student knowledge to select an algorithm design paradigm using multiple-choice questions. We present the construction of the \textit{algorithmic paradigm selection assessment} (APSA) and preliminary data demonstrating its effectiveness as an assessment. We discuss the key points we learned during this process to write multiple-choice questions for Algorithm Design Paradigms. We tested the internal consistency of our assessment using Cronbach's $\alpha$ and obtained a score of $0.73$, which is above the required threshold of $0.7$. APSA can be used across institutions as a standardized way to assess students' ability to select different algorithm design paradigms. APSA will assist researchers in evaluating whether a theory helps students improve their knowledge of different Algorithm Design Paradigms.
\end{abstract}

\ccsdesc[500]{Social and professional topics~Computing education}
\ccsdesc[500]{Applied computing~Education}
\ccsdesc[500]{Social and professional topics~Student assessment}


\keywords{
Algorithmic Paradigm,
Algorithm Design Problem,
Assessment
}


\maketitle

\section{Introduction}

The \textit{algorithmic paradigm selection assessment} (APSA), presented in this work, measures students' ability to select the appropriate algorithmic paradigm for a given \textit{algorithm design problem}. Algorithm design problems are generally covered in upper-level algorithms courses. They have well-defined inputs and outputs, for which a solution is a short amount of code or pseudocode. To solve these kinds of algorithm design problems, students need to learn high-level algorithmic design paradigms. In this paper, we focused on design problems for four algorithm design paradigms taught in algorithm courses at the undergraduate level ~\citep{acm2013curriculum,luu2023algorithms}: dynamic programming (DP), divide-and-conquer (D\&C), greedy algorithms (GA), and graph modeling (GM), where the problems are solved using standard graph algorithms such as path-finding algorithms, breadth-first search, and depth-first search. 

ACM curriculum emphasizes teaching computer science students to identify algorithm paradigms for a design problem~\cite{acm2013curriculum}:
\begin{itemize}[leftmargin=1em, rightmargin=1em, label={}, itemsep=0.1em, parsep=0.4em]
   \item \emph{``An important part of computing is the ability to select algorithms appropriate to particular purposes and to apply them,
 recognizing the possibility that no suitable algorithm may exist.''}
\end{itemize}

However, existing research has shown that students struggle to figure out which algorithmic paradigm to use for a problem, even if they have learned the correct on for the problem at hand~\cite{zehra2018student, shindler2022student, armoni2009reduction, ginat2004senior}. Being able to effectively test students' ability to identify the correct algorithmic paradigm at scale is the first step toward creating and evaluating interventions to help teach this skill to students.

In this paper, we present the APSA, an assessment that evaluates students' knowledge through a set of multiple-choice questions. The researcher will be able to assess a large student population in a limited timeframe and with less manual labor. APSA could assist instructors in assessing students' ability to implement algorithms for a given design problem and in making  modifications to the teaching strategy to help students better understand the subject matter.

Here, we present the steps for generating questions for APSA, validating their quality, and what we learn from them. We want to answer the following research questions in this paper:
\begin{itemize}
\item[\textbf{RQ1:}] What guidelines do we need to follow to write assessment questions to evaluate students' ability to select an algorithm design paradigm?
\item[\textbf{RQ2:}] What does the statistical evidence say about the reliability of the APSA and validity of questions?
\end{itemize}

We administered the APSA to students from two public universities in the USA and used Cronbach's $\alpha$ to measure its reliability.
Through the process, we developed guidelines for writing questions to assess students' selection of algorithmic paradigms effectively, and we were able to construct a set of questions that work effectively for this purpose. Questions will be published under a Creative Commons license with the final version of the paper (omitted now for anonymous review). 

\begin{table}[ht]
    \caption{Distribution of number of questions over algorithm paradigm.}
    \centering
    \begin{tabular}{c|c|c}
         & Version $1$ & Version $2$ \\\hline
        Dynamic Programming & 8 & 9 \\
        Divide-and-Conquer & 8 & 8 \\
        Graph Modeling & 6 & 5 \\
        Multiple Options & 8 & 6 \\\hline
        Total & 30 & 28 \\\hline
    \end{tabular}
    \label{tab: question_distribution}
\end{table}

\section{Related Work}{\label{sec: rel_work}}
\subsection{Algorithms Education}

While students begin learning about algorithms in their introductory computing sequences, most computer science degrees include a mid- to upper-level course dedicated solely to the study of algorithms, as suggested by the ACM curricular guidelines~\cite{acm2013curriculum,kumar2024computer}.
These courses typically focus on algorithm design techniques such as dynamic programming, divide-and-conquer, and graph modeling algorithms~\cite{luu2023algorithms}. Even after learning algorithmic techniques, students often fail to recognize which one to apply~\cite{zehra2018student, shindler2022student, armoni2009reduction, ginat2004senior}.

\citet{ginat2004senior} and \citet{armoni2009reduction} found that students were unable to apply their knowledge to use recursion and reduction effectively.
\citet{zehra2018student}, \citet{shindler2022student}, and \citet{liu2025student} all interviewed students solving dynamic programming problems. They all found that students use the divide-and-conquer and greedy approaches when solving the given dynamic programming problem. 
\citet{chen2025novice} interviewed students solving graph layering problems, and similarly found that students would try to use dynamic programming and greedy approaches when they were not appropriate.



While all of this research focuses on student behavior and misconceptions 
of students while working on problems for a particular algorithmic paradigm, we instead focus on the assessing student ability to pick between design paradigms. The first work to explicitly examine the process of students attempting to select which algorithmic paradigm to use was~\citet{pradhan2026students}, who interviewed students working on divide-and-conquer, dynamic programming, and graph modeling questions. Similar to prior work, they found that students frequently failed to identify the correct algorithmic paradigm to use for algorithm design problems, regardless of which algorithmic paradigm was the correct choice for the problem. They also proposed a systematic way of teaching students how to select an algorithmic paradigm, though they did not yet provide evidence for its effectiveness. The validation of the APSA, which we present in this paper, represents the first step toward being able to empirical test the effectiveness of methods for teaching student how to select an algorithm design technique.


\subsection{Validated Measurement Instruments in CS Education}
As measurement of various phenomena is inherent to education research, many validated 
assessments have been created for computer science~\cite{decker2019topical}, and many validated instruments created outside of computer science education have been used by the computer science education research community~\cite{margulieux_review_2019}.
Most closely related to the current work,~\citet{ferland2025construction} have created a concept inventory for dynamic programming, and~\citet{farghally2017towards} created a concept inventory for algorithm analysis topics. \citet{danielsiek2017instrument} created a survey to assess student's algorithmic self-efficacy.
However, there is no validated instrument for testing student's ability to select an algorithmic paradigm. In fact, the authors are not aware of any attempt to measure paradigm selection through autogradable questions.
Though it is a cognitive assessment (rather than a non-cognitive one such as those around self-efficacy and related phenomena)~\cite{decker2019topical}, we do not consider the APSA to be a concept inventory because it focuses on an exact technical skill---that of being able to select the appropriate algorithmic paradigm for a given problem---rather than focusing on conceptual knowledge as concept inventories do.

\section{Methods}{\label{sec: val_method}}

\begin{figure*}[ht]
    \centering  \includegraphics[width=0.7\linewidth]{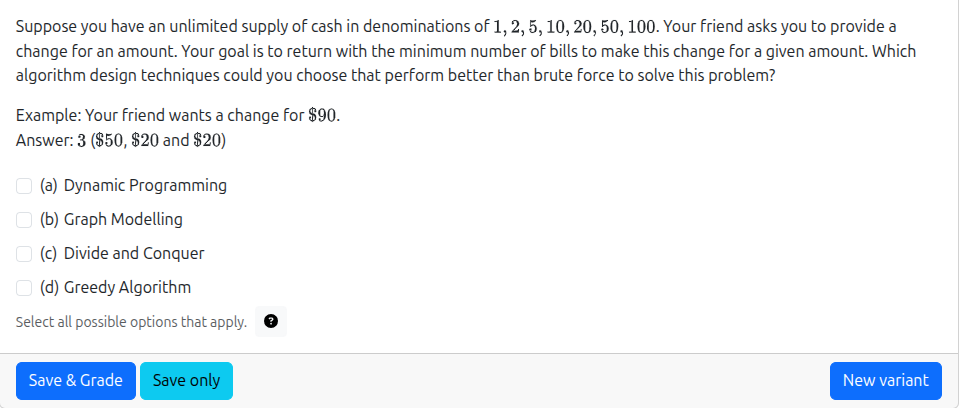}
    \caption{Example of the questions given to the student. Each question has a problem statement with an example and four options in a fixed order. Users have to choose only one option in Example 1 while having to select multiple options in Example 2. We used PrairieLearn to test students.}
    \label{fig: example}
\end{figure*}

\subsection{Question Generation}\label{ques_generation}
The first author drafted the multiple-choice questions to evaluate students' knowledge to choose a suitable algorithm design paradigm for a given algorithm design problem. The second and third authors, who both have years of experience teaching algorithms courses, helped in revising the questions.

We reviewed multiple problem-example sources, such as books~\cite{erickson2023algorithms, sedgewick2011algorithms, cormen2022introduction, chinmoy2016cracking}, LeetCode~\cite{leetcode}, and GeeksforGeeks~\cite{geeksforgeeks} to get inspiration on how to write problems. Then we generated multiple-choice questions using our own wording and scenarios to create a new problem for our study. 

We wrote all questions in a standardized format: students were given an algorithm design problem with an example to demonstrate, and they had four options: Dynamic Programming, Graph Modeling, Divide-and-Conquer, and Greedy Algorithms, as shown in Figure~\ref{fig: example}. The four options were in the same order in each question for simplicity. We collected assessment data in two versions (Version~$1$ and Version~$2$) of the APSA question set, which evaluates students' ability to select the correct algorithm design paradigm. 

Many validated assessments have between 20 and 30 questions, as this strikes a good balance between obtaining accurate measures of student knowledge and not taking too much time to administer~\cite{murtaza2023taking}. We had different numbers of questions in the assessment across versions. There were a total of $30$ multiple-choice questions in Version~$1$ covering the diverse use cases of algorithm design paradigms. 

After analyzing the results from Version~$1$, we re-evaluated the questions and assessment criteria for Version~$2$ of the study. We expected that we would have to throw out some questions from Version $1$, so we started with $30$ questions; however, for Version~$2$, we only had $28$ questions, as we knew we would not have to throw out as many questions in Version $2$ since we had used the findings from Version~$1$ for question generation. $16$ questions of Version~$2$ were from Version~$1$ while we wrote $12$ new questions. 

The distribution of questions in Versions~$1$ and $2$ is as shown in the Table~\ref{tab: question_distribution}. Of the $8$ questions with multiple options in Version~$1$, only $4$ had Greedy Algorithm as one of the answers, and reused $3$ in Version~$2$. 
We decided to deemphasize Greedy Algorithms in the APSA, because situations where they work are rare, and thus many popular algorithms books and instructors guide students away from using them in general~\cite{halim2018competitive,erickson2023algorithms}. 

For questions where multiple algorithm paradigms could work, in  Version~$1$, we gave students credit for picking any one of the paradigms that was possibly correct. However, this resulted in those questions being too easy to give an effective measurement of student knowledge (as shown in Table~\ref{tab: result}). 
To resolve this, we used a more nuanced approach in Version~$2$, wording the questions differently for each of these three cases:

\begin{enumerate}
	\item Single Algorithm Design Paradigm: Which algorithm design technique would you choose that performs better than brute force to solve this problem?
	\item Combine Multiple Design Paradigms: Which algorithm design techniques should you use that perform better than brute force to solve this problem?
	\item Multiple Algorithm Design Solutions: Which algorithm design techniques could you choose that perform better than brute force to solve this problem?
\end{enumerate}

Where for types (2) and (3), the student had to select all correct answers in order for their solution to be marked correct.

\subsection{Participants}
For Version $1$, $28$ students participated in our test; they were all students at \usu~ and took the Algorithm Design course in Fall 2024. All $28$ students completed the assessment. In Version $2$, out of $393$ students from \uci~ enrolled in the Algorithm Design course in Spring 2025, $304$ students completed the test in the second round. 

In \usu~, the semester lasted four months, during which the instructor taught Dynamic Programming for four weeks, Graph Modeling for three weeks, Divide-and-Conquer for two weeks, and Greedy Algorithm for one week. Similarly, the term at \uci~ lasted three months, and the percentage of course content was similar. At both universities, the instructors followed similar course material and administered the APSA test to students as a take-home exam with a one-week deadline. Each student who attempted all the APSA questions was given extra credit for the course. In both cases, the test was administered shortly before the final exams, so students knew about all four algorithm design paradigms needed for the test.
We collected the data with the authorized IRB from both universities. We used \pl{} to collect assessment scores.


\subsection{Analysis}
To evaluate the APSA, we used Classical Test Theory (CTT), a measurement tool for assessing test reliability~\cite {devellis2006classical}, which is often valuable for smaller datasets and for initial evaluation of assessments~\cite{jorion2015analytic, offenberger2019initial, ferland2025construction}. CTT calculates Cronbach's $\alpha$ to assess the reliability of APSA. Cronbach's $\alpha$ measures the correlation between questions to measure their consistency~\cite{cronbach1951coefficient}. Cronbach's $\alpha$ greater than $0.8$ is considered good, while $0.7$ is the threshold value to be considered satisfactory~\cite{offenberger2019initial}. While the second-round data collection was large enough to employ item response theory (IRT), we continued to use CTT to enable comparisons of the statistics between versions. In most real-world data, when measuring difficulty and discrimination, CTT performs as well as IRT~\cite {fan1998item}.


To evaluate the effectiveness of questions in APSA, we used the metrics \textit{difficulty} and \textit{discrimination}. Difficulty is the proportion of students who answer a question correctly, and a value from $0.2$ to $0.8$ is considered ideal~\cite{offenberger2019initial}. If most students get similar scores, the test is less effective, so using a range of $0.2$ to $0.8$ reduces the ceiling and floor effects of the test. A lower difficulty value means the question is more challenging, while a value close to 1 suggests the student finds the question easy. Meanwhile, discrimination is the point-biserial correlation between a question score and total test score~\cite{offenberger2019initial}. For example, a negative discrimination score for a question indicates that students who did better on the exam as a whole did poorly on that question.

In contrast, a high positive discrimination indicates a strong correlation between students' scores on that question and their overall scores. A discrimination value greater than $0.2$ is considered acceptable. We have labeled a question with suitable difficulty scores and discrimination greater than $0.2$ as a \textit{good} question in this paper. For any other case, we label the question as \textit{poor}.

\section{Results}{\label{sec: val_results}}
\begin{table}
\centering
\caption{Difficulty (Diff.) and Discrimination (Disc.) scores of questions in Version 1 and Version 2 of the study. The algorithm column denotes the algorithm design paradigm used to solve each algorithm design problem. DP denotes Dynamic Programming, DC is Divide-and-Conquer, GA is Greedy Algorithm, and GM is Graph Modeling. Questions with multiple solutions have combined names: ``and'' indicates the combination of design paradigms for a solution, and ``or'' represents any one of the given paradigms that can solve the problem.}
\begin{tabular}{|c|cc|cc|c|}
    \hline
    & \multicolumn{2}{c|}{Version $1$} & \multicolumn{2}{c|}{Version $2$} &\\\hline
    Ques. & Diff. & Disc. & Diff. & Disc. & Paradigm \\\hline
    4 & 0.89 & 0.48 & & & DP\\
    8 & 0.93 & 0.43 & & & DP\\
    15 & 0.89 & 0.40 & & & DP\\
    19 & 0.79 & 0.44 & 0.70 & 0.38 & DP\\
    20 & 0.75 & 0.53 & 0.67 & 0.46 & DP\\
    22 & 0.39 & 0.42 & 0.43 & 0.23 & DP\\
    25 & 0.79 & 0.04 & & & DP\\ 
    27 & 0.61 & 0.22 & 0.55 & 0.36 & DP\\
    31 & & & 0.53 & 0.35 & DP\\
    32 & & & 0.56 & 0.39 & DP\\
    34 & & & 0.44 & 0.39 & DP\\
    38 & & & 0.69 & 0.33 & DP\\
    40 & & & 0.76 & 0.43 & DP\\\hline\hline
    5 & 0.50 & 0.30 & & & D\&C\\
    10 & 0.64 & 0.03 & & & D\&C\\
    13 & 0.79 & 0.18 & 0.72 & 0.41 & D\&C\\
    18 & 0.79 & 0.47 & 0.79 & 0.45 & D\&C\\
    23 & 0.46 & 0.22 & 0.30 & 0.17 & D\&C\\
    26 & 0.68 & 0.42 & 0.57 & 0.34 & D\&C\\
    29 & 0.75 & 0.34 & 0.70 & 0.39 & D\&C\\
    30 & 0.86 & 0.46 & & & D\&C\\
    35 & & & 0.66 & 0.45 & D\&C\\
    37 & & & 0.33 & 0.24 & D\&C\\
    39 & & & 0.77 & 0.45 & D\&C\\\hline\hline
    6 & 0.61 & 0.15 & & & GM\\
    9 & 0.71 & 0.51 & 0.48 & 0.31 & GM\\
    12 & 0.46 & 0.59 & 0.49 & 0.34 & GM\\
    16 & 0.64 & 0.32 & 0.45 & 0.28 & GM\\
    21 & 0.68 & 0.10 & & & GM\\
    28 & 0.96 & -0.05 & & & GM\\
    36 & & & 0.65 &	0.47 & GM\\
    41 & & & 0.66 & 0.34 & GM\\\hline\hline
    1 & 0.96 & 0.20 & 0.3 & 0.27 & GA and GM \\ 
    17 & 0.96 & 0.33 & & & GA and GM \\ \hline\hline
    11 & 0.36 & -0.13 & & & GA or GM\\
    7 & 0.93 & -0.03 & & & GA or GM \\\hline\hline
    2 & 0.89 & 0.33 & 0.41 & 0.42 & DP or GA\\
    3 & 0.71 & 0.1 & 0.42 & 0.33 & DP or GA \\\hline\hline 
    14 & 0.93 & -0.12 & & & DP or GM \\
    24 & 0.89 & -0.01 & 0.47 & 0.25 & DP or GM\\
    33 & & & 0.34 & 0.38 & DP or GM\\\hline\hline
    42 & & & 0.33 & 0.27 & DP and GM\\\hline
\end{tabular}
\label{tab: result}
\end{table}

\begin{table}[ht]
    \centering
    \caption{ An example of reworded longest common sequence question used in APSA and similar question shown in class. The question on the APSA was too easy for students who had seen the original, despite having been reworded.}
    \begin{tabular}{|p{4cm}|p{4cm}|}
    \hline
    Question student saw in class & Reworded question in APSA \\\hline
    A subsequence is a sequence contained within another sequence. Example subsequence: ‘AAA’, ‘BBB’, and ‘AB’ and all subsequences of ‘ABABAB'.
    Note: a subsequence is not the same as a substring. 
    The longest common subsequence (LCS) of two sequences is the longest sequence that is a subsequence of both. 
    
    Example LCS: The LCS of “blueberry” and “boysenberry” is “beberry”, 
    LCS length: 7 
    &
    Suppose you are given a list of positive numbers. Find the longest subsequence in the list that is strictly increasing. The sublist does not have to be continuous. Which algorithm design technique would you choose that performs better than Brute force to solve this problem?
    
    For example: nums = [1, 5, 15, 3, 35, 2, 20]
    Answer: [1, 5, 15, 35] or [1, 5, 15, 20] \\\hline
    \end{tabular}
    \label{tab: examplesimilarquestion}
\end{table}


Table ~\ref{tab: result} shows the difficulty and discrimination score of questions from Version $1$ and Version $2$ of the APSA.

\subsection{Version 1}
In Version $1$, most of the questions had difficulty scores between $0.39$ and $0.96$. We had $11$ questions with difficulty scores greater than $0.8$, which typically indicate that students found them easy. The ideal difficulty range is $[0.2, 0.8]$. We had five questions with negative discrimination scores, which shows students who performed well in overall exams had poor performance on these questions. We also had $6$ questions whose discrimination scores were below the ideal bound of $0.2$. In Version $1$ of the question validation, we had only $13$ good questions as shown in Table ~\ref{tab: result}.

\begin{figure}[ht]
  \centering  \includegraphics[width=0.9\linewidth]{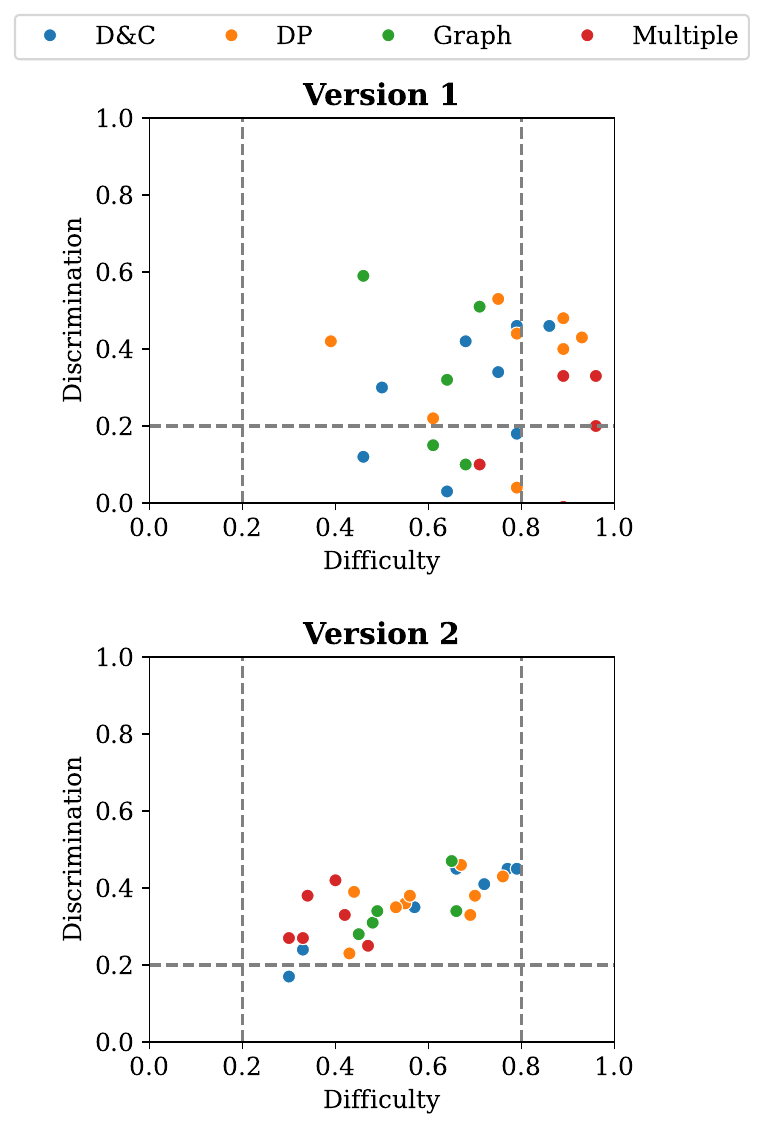}
  \caption{Difficulty and discrimination scores of questions from Version 1 and Version 2. The dotted line at 0.2 and 0.8 on the x-axis shows the range for the difficulty score, while the discrimination threshold is 0.2. Questions close to Difficulty 1 are easy, while Questions close to Discrimination 1 mean the high scorer performs better in the question. Version 2 questions were more reliable than Version 1. The bold circle shows that more than one question has the same Difficulty and Discrimination score.}
  \label{fig: question_details}
\end{figure}


\subsection{Modifications for Version 2}
Since the Cronbach's $\alpha$ from the Version $1$ was less than the threshold, we made modifications to some questions, and replaced other questions entirely for Version~$2$ of the APSA. We used $12$ good questions from Version $1$, those in the desired range of difficulty and discrimination scores. We made a mistake and failed to keep $Q5$ in Version $2$. When analyzing Version $1$, we noticed that questions that could be solved using multiple design paradigms had poor difficulty and discrimination, as shown in Table~\ref {tab: result}. We only asked students to select one option from numerous possible solutions, and therefore, it increased the probability of picking the correct answer for the problem.

Dynamic Programming is considered one of the most demanding algorithm design techniques to master~\cite{enstrom2017iteratively, shindler2022student}. However, Version $1$ data shows students found it easy to identify DP problems. Out of $8$ problems that had only DP as a solution, $3$ had difficulty above $0.8$ (Q$4$, Q$8$, and Q$15$). The three questions that had difficulty above $0.8$ were longest common sequence (Q$4$), coin-row (Q$15$),  and Fibonacci series (Q$8$). Even though we reworded the question to make it different than the standard problem in the study materials, they were able to identify the correct algorithm design paradigm with ease. An example of this is shown in Table ~\ref{tab: examplesimilarquestion}. So, for Version $2$ we focused on writing dynamic programming problems that were entirely different from the examples commonly used in classes. 

For Graph Modeling problems, using keywords like path, travel, and network helped students to choose their preferred algorithm design paradigm. When we avoided such keywords from the Graph Modeling problems, the questions performed better with difficulty scores between $0.2$ and $0.8$. Also, we found that very few students chose the Greedy solution for the given problem. Previous research~\cite {shindler2022student, ginat2003greedy} showed students used greedy as a part of the solution even if the approach is incorrect.
However, we see in $6$ questions where greedy is a solution, students tend to select the other option as their primary solution. The instructors from both universities taught students not to use greedy, unless they are really sure it is correct. ~\citet{erickson2023algorithms} also discourages students from using greedy.

One of our research questions is to identify the patterns for successfully writing APSA questions. From our study, we found that following these three steps helps us write these problems:
\begin{enumerate}
      \item In Version~$1$, we asked the student to select only one answer, even when multiple answers were possible. These problems did not perform well and were less difficult. However, in Version~$2$, the students had to select all possible design paradigms, which made the problem challenging, as shown by the difficulty matrix score in Table~\ref{tab: result} for questions Q$1$, Q$2$, Q$3$, and Q$24$. We should ask students to select all possible algorithm design paradigms to effectively test their knowledge of algorithms, as shown in Figure~\ref{fig: example}.
    \item In Version~$1$, we tried rewording the problems students have seen in class and used different variations of those questions; however, students were able to answer those problem correctly. These questions were not sufficient to test the students properly. We should focus on designing a problem not covered in class lectures or homework.
    \item Try to avoid keywords such as \textit{path}, \textit{travel}, and \textit{network} while writing a new question for Graph Modeling design problems. We generated new problems that use the concept of a graph but omit keywords such as \textit{path} and \textit{network}. We worded the question Q$28$ in Version~$1$ as follows:
    \begin{quote}You need to travel from point A to point B, but external factors might delay the estimated arrival time. The goal is to make the trip in the minimum time. Which algorithm design technique would you choose that performs better than Brute force to solve this problem?\end{quote}
    We used keywords such as travel and minimum time in Q$28$, which indicated the problem as a graph, so we only got a difficulty score of $0.96$. However, in Version $2$, we did not use any such keyword, so we got a preferred discrimination and difficulty score.
\end{enumerate}

\subsection{Version 2}
After making the required modifications as described in~\ref{ques_generation}, we reassessed the quality of the questions with students from \uci. From Table~\ref{tab: result}, we can see that the difficulty score of all questions is between $0.2$ and $0.8$, which is considered ideal. The variation in difficulty score suggests we have questions of different difficulty to assess students' mastery of the algorithm design paradigm. Similarly, the discrimination score was greater than $0.2$ for most questions, with the exception of $Q23$ ($0.17$). In the exception of this one question, the discrimination shows that students who get good scores in overall exams perform better in these questions. These discrimination parameters mean that the question do accurately measure student knowledge.

Figure~\ref{fig: question_details} visualizes the validity of the questions from Version $1$ and Version $2$. Version $2$ has most of the questions in the reliable zone, where $0.2 \leq \textit{difficulty} \leq 0.8$ and $\textit{discrimination} \geq 0.2$, whereas many of the questions in Version $1$ were too easy and were therefore less reliable. In both versions, most students were able to complete all questions in under 30 minutes.


\begin{table}[ht]
  \centering
  \caption{Cronbach's $\alpha$ score after Version 1 and Version 2 data collections. In Version 2, we were able to achieve above $0.7$, commonly believed to be satisfactory for assessments.}
  \begin{tabular}{|c|c|}
    \hline
    Versions & Cronbach's $\alpha$ \\\hline\hline
    Version~$1$ & 0.49 \\\hline
    Version~$2$ & 0.73 \\\hline
    Version~$2$ - Balanced Question Set & 0.72\\
    \hline
    \end{tabular}
  \label{tab: cronbach}
\end{table}


We measured the reliability of APSA using Cronbach's $\alpha$, which is also our RQ2. Cronbach's $\alpha$ is a reliability coefficient that helps assess the assessment. Cronbach's $\alpha$ indicates that the questions are reliable and measure a good understanding of the algorithm design paradigm. We calculated Cronbach's $\alpha$ of the APSA in Versions~$1$ and~$2$. The Cronbach's $\alpha$ calculated from Version~$1$ is $0.49$ and $0.73$ for Version~$2$ as shown in Table~\ref{tab: cronbach}. In Version~$2$, we obtained a satisfactory threshold value of $0.7$ for an internally consistent assessment. In Version~$1$, we had only $13$ good questions (Table~\ref{tab: result}), so calculating Cronbach's $\alpha$ for only the good questions yields an improved $\alpha$ of $0.63$. Similarly, in Version $2$, we obtain the Cronbach's $\alpha$ value of $0.74$ after removing $Q23$ (since the discrimination score is less than 0.2) as shown in Table~\ref{tab: result}, which is greater than the reliability threshold value.

\begin{table}[h]
\caption{List of questions in Version 2 of APSA.  DP denotes Dynamic Programming, DC is Divide-and-Conquer, GA is Greedy Algorithm, and GM is Graph Modeling. ``and'' indicates the combination of design paradigms for a solution, and ``or'' represents any one of the given paradigms that can solve the problem.}
\begin{tabular}{|c|p{3.02cm}|c|}
    \hline
    Algorithm Design Paradigms & Questions & Total\\\hline
    DP & Q19, Q20, Q22, Q27, Q31, Q32, Q34, Q38, Q40 & 9 \\\hline
    D\&C & Q13, Q18, Q23, Q26, Q29, Q35, Q37, Q39 & 8 \\\hline
    GM & Q9, Q12, Q16, Q36, Q41 & 5 \\\hline
    GM and GA & Q1 & 1 \\\hline
    DP or GA & Q2, Q3 & 2 \\\hline
    DP or GM & Q24, Q33 & 2 \\\hline
    DP and GM & Q42 & 1 \\\hline
\end{tabular}
\label{tab: questionV2}
\end{table}

\subsubsection{Balancing the APSA Across Design Paradigms}
Version $2$ is unbalanced, with more questions have Dynamic Programming (DP) as the correct algorithm design paradigm than any other paradigm, as shown in Table~\ref{tab: questionV2}. Because most DP problems in Version $1$ for did not perform well, we wrote more questions with DP for Version $2$ in anticipation of needing to throw out some of them for poor performance again. However, for the second round, all DP questions had acceptable difficulty and discrimination scores.

If someone wants to use a version of the APSA with a more balanced number of questions across the different algorithmic paradigms, we recommend removing Q$3$, Q$22$, Q$27$, and Q$40$. After removing these questions, we will have a total of $24$ questions and the difficulty of the questions will be roughly uniformly distributed, and Cronbach's $\alpha$ is $0.72$, which is still above the required threshold as shown in Table~\ref{tab: cronbach}. This modification gives an exam with 6 DP-only questions, 8 D\&C-only questions, 5 GM-only questions, 1 GA and GM questions, 1 DP and GA question, 3 DP and/or GM questions, for a total of 10 DP, 8 D\&C, 9 GM, and 2 GA questions. 

We will make the questions from Version $2$ publicly available after the review process.

\section{Discussion}{\label{sec: discussion}}
Many questions did not perform well in the first version of APSA. However, that version helped us to learn different things to consider while writing questions. We reused the \textit{good} questions from Version~$1$ as well as added new questions for Version~$2$. The resulting set, evaluated in Version~$2$, was largely reliable when measuring students' knowledge of algorithm design paradigm selection. The questions used in Version~$2$ covered the paradigms of algorithm design that are primarily taught in undergraduate education~\citep{luu2023algorithms}: Dynamic Programming, Divide-and-Conquer, Greedy Algorithm, and Graph Modeling, as shown in Table~\ref{tab: questionV2}. We measure the validity of each question to assess the students' knowledge in selecting the right algorithm design paradigm for each given problem.

Previously, researchers used interviews with students to evaluate their knowledge of Algorithm Paradigms. However, students could only solve two or three questions in an hour-long interview~\cite{shindler2022student, 10.1145/3641554.3701888}. The APSA could be useful to evaluate teaching interventions during randomized controlled trials, when time to assess student knowledge on pre- and post-tests is limited.

\section{Limitations and Future Work}{\label{sec: lim_future}}
One of the limitations we have is that we only tested the APSA in two universities, and so testing with a more broad population would give further confidence in its generality. Also, our current validation does not focus on the assessment testing of students from different demographics.

In the future, we can conduct validation studies with populations across more universities, and with collecting demographic information. We also plan to conduct controlled experiments to test different instructional methods, to find the best instructional method for helping students in selecting an algorithmic paradigm.


\section{Conclusion}{\label{sec: conclusion}}
In this paper, we discussed the detailed process of writing questions for \textit{algorithmic paradigm selection assessment} (APSA). We did not obtain Cronbach's $\alpha$ above the required threshold for the Version $1$ assessment of the question set with students from \usu. However, after making the necessary changes to the question set, we obtained a Cronbach's $\alpha$ of $0.73$, which exceeds the required reliability threshold when tested with students from \uci. We were able to write and validate multiple-choice questions that could be useful to evaluate students' knowledge in selecting the correct algorithmic paradigm. The APSA will serve as a benchmark to help researchers and educators more effectively teach students the algorithmic paradigm selection process.

\bibliographystyle{ACM-Reference-Format}
\bibliography{references}
\end{document}